\documentclass[12pt,preprint]{aastex}

\shorttitle{Vector magnetic fields of Solar Granulation}
\shortauthors{Jin et al.}

\begin{document}


\title{Vector magnetic fields of Solar Granulation}

\author{Chunlan Jin, Jingxiu Wang, and Meng Zhao}

\altaffiltext{}{National Astronomical Observatories, Chinese
Academy of Sciences, Beijing 100012, China; E-mail:
jinchunlan@ourstar.bao.ac.cn, wangjx@ourstar.bao.ac.cn,
zhaomeng@ourstar.bao.ac.cn}

\begin{abstract}
Observations of quiet Sun from the Solar Optical
Telescope/Spectro-Polarimeter (SOT/SP) aboard the \emph{Hinode}
spacecraft would reveal the magnetic characters of the solar
photosphere. By making use of the deep mode observations of three
quiet regions, we have statistically studied the vector magnetic
fields of solar granulation. More than 2000 normal granules are
manually selected to form a sample. It is recognized that some
granules are even darker than the mean photosphere in intensity,
and there is a linear correlation between intensity and Doppler
velocity in granules.

The distributions of longitudinal and transverse apparent magnetic
flux densities, Doppler velocity and continuum intensity of
granules are obtained, and their unsigned magnetic flux measured.
Two approaches are carried out in this study. First we obtained
the magnetic properties of granulation by averaging the
measurements for all the sampling granules. Secondly, we
reconstructed an average granular cell based on a sub-sample, and
obtained the detailed distribution of apparent magnetic flux
density within the model granular cell. All the results have been
compared with that of inter-granular lanes and a few typical
abnormal granules.

Our statistic analysis reveals the following results: (1) The
unsigned magnetic flux of individual granule spans the range from
$1.1\times 10^{15}$ Mx to $3.3\times 10^{18}$ Mx with a peak
distribution at $1.6\times 10^{16}$ Mx; (2) The unsigned
longitudinal apparent flux density of granules ranges from almost
zero to 212 Mx $cm^{-2}$ with a mean longitudinal apparent flux
density of 12 Mx $cm^{-2}$, while the transverse apparent flux
density of granules ranges from 4 to 218 Mx $cm^{-2}$ with a mean
transverse apparent flux density of 79 Mx $cm^{-2}$. The
longitudinal and transverse apparent magnetic flux densities of
granules are positively correlated, and the longitudinal apparent
flux density of granules is weaker than the corresponding
transverse apparent flux density; (3) The magnetic inclination of
granules with respect to the surface that's perpendicular to the
line-of-sight falls in the range of 4.8 to 76.7 degrees with a
peak distribution at 25 degrees. On average, the magnetic vectors
in granules are more vertical than that in the inter-granular
lanes; (4) There is a strong preference that both the vertical and
horizontal fields on the quiet Sun reside in the inter-granular
lanes; (5) The detailed distributions of apparent flux density,
Doppler velocity, and continuum intensity within an average
granular cell are presented. These distributions can well be
empirically formulated.

\end{abstract}

\keywords{Sun: granulation --- Sun: magnetic fields --- Sun:
photosphere --- techniques: polarimetric}

\section{Introduction}

Under good conditions of atmospheric seeing, the observations of
the solar surface with a telescope of at least 30 cm aperture can
reveal the cellular pattern that covers the entire solar surface,
except the sunspots region; this pattern is called granulation.

As the smallest convective elements, many properties of granules
have been carefully studied. The observed size of granules
excluding the surrounding dark lanes ranges from approximately
0$''$.2 (the limit of ground-based observations) to approximately
3$''$.4 (Peter N Brandt 2000), and the mean granular size amounts
to 1$''$.1 (Namba \& Diemel 1969), or 1$''$.35 (Bray et al. 1984).
The mean cell size of the granular elements including one-half of
the surrounding dark lanes is 1$''$.94 according to Bray \&
Loughhead (1977), and 1$''$.76 according to Roudier \& Muller
(1986). The granulation is a non-stationary phenomenon. Frequently
the granules expand and split into smaller components that drift
apart, and the fragments may turn grow and fragment, or merge with
others, or shrink and decompose. The time for the granular
photospheric intensity decay to $1/e$ of an initial value is about
6 minutes according to a number of studies (e.g., Bahng $\&$
Schwarzschild 1961; Title et al. 1986). Individual granules may
live longer than this, e.g., over 8 minutes as identified by
Mehltretter (1978), and a mean lifetime of 12 minutes found by
Dialetis et al. (1986). An important feature of granular evolution
is the proper motion of granules. From the high-resolution slit
spectrograms taken at disk center, one can measure the vertical
velocity related to granular motion, and the peak value identified
is in the range 1.5-2.0 km/s, after removing the velocity
components due to oscillations. Furthermore, the horizontal
velocity components are larger than the vertical components by a
factor of 2 (Mattig et al. 1981).

The granular motions also seem to have significant influence on
the magnetic emergence. The horizontal inter-network fields
reported by Lites et al. (1996), with typical sizes of 1$''$ and
lifetime of $~$5 minutes, suggest that small magnetic loops are
being advected toward the surface by the upward motion of the
plasma inside the granules. If the horizontal inter-network fields
are indeed emerging flux, Lites et al. (1996) estimated that the
rate of magnetic flux driven to the surface by the upward motion
of the plasma inside the granules is greater than the rate of flux
emergence in bipolar sunspot regions averaged over the whole solar
cycle. The horizontal motions inside the granules carry the
vertical magnetic flux toward the inter-granular lanes (Harvey et
al. 2007; Centeno et al. 2007).

Furthermore, the magnetic emergence also has important influence
on the shape of the underlying granulation pattern. The magnetic
emergence of flux tubes with a flux of more than $10^{19}$ Mx of
longitudinal flux disturbs the granulation and leads to the
transient appearance of a dark lane (Cheung et al. 2007), that is
so-called `abnormal granulation'. The abnormal granulation is also
a term sometimes used for granulation in plage regions. While
small-scale flux tubes with less than $10^{18}$ Mx are not
sufficiently buoyant to rise coherently against the granulation
and produce no visible disturbance in the granules (Cheung et al.
2007). Therefore, one of the crucial points to understand the
structure of the granulation is how the convective elements are
modified by the presence of surface magnetic fields.

In the photospheric layer, the plasma $\beta$ is close to 1,
thusly, the magnetic fields and the convection of plasma in the
photosphere are vigorously interacting all the time. To quantify
the magnetic properties of solar granulation will provide clues in
understanding the interaction, and test the theories and numerical
simulations of this magnetic convection.

Unfortunately, limited by the spatial resolution and sensitivity
of polarization measurements, for this very fundamental convection
cells, their magnetic properties, i.e., the magnitude and
distribution of magnetic fields in granules are poorly known. With
infrared polarimetry Lin \& Rimmele (1999) identified magnetic
elements with flux less than $5\times 10^{16}$ Mx and field
strength of 200 - 1,000 G, which they called \emph{granular}
magnetic fields. The spatial distribution and time evolution of
these magnetic features are closely associated with the solar
granulation. These magnetic elements seem to occupy 68\% of the
quiet Sun area. From wavelength-integrated measures of
\emph{Hinode} Spectro-Polarimeter (SP), Lites et al. (2008)
uncovered the horizontal magnetic flux on the quiet Sun. They
obtained average horizontal and vertical flux density in quiet
inter-network regions, which were 55 Mx cm$^{-2}$ and 11 Mx
cm$^{-2}$, respectively. These authors found evidence that
magnetic fields are organized on mesogranular scales, with both
horizontal and vertical fields showing ¡®voids¡¯ of reduced flux
density of a few granules spatial extent. Moreover, the vertical
fields are concentrated in the inter-granular lanes, whereas the
stronger horizontal fields occur most commonly at the edges of the
bright granules.

The unprecedented spatial resolution and sensitivity of the Stokes
Spectro-Polarimeter onboard of \emph{Hinode} seem to provide a new
opportunity to advance our knowledge on the interaction between
the magnetic field and granulation. A very key question is whether
or not the magnetic field in inter-network regions can be modified
and organized by granulation, the smallest convection elements on
the quiet Sun. Getting the average properties of magnetic
characters in granules and their overall correlation with
convection flow and intensity become the first step of the
relevant studies and serves as the aim of this current work. With
this goal we analyze the the \emph{Hinode} Spectro-Polarimeter
data inverted by the full Stokes profiles based on the assumption
of Milne-Eddington atmosphere model, and make two simple
approaches: to get longitudinal and transverse magnetic flux
densities, convection flows, and intensity of all the 2,100
sampling granules and to obtain the detailed distributions of
magnetic flux densities within a model granular cell.

In Section 2, we describe the observations used for this study and
the strategy of Stokes profile inversion. In Section 3, we obtain
the statistical properties of granular vector magnetic fields and
other relevant parameters. In Section 4, we put forward a model of
average solar granular cell, and get detailed distributions of
magnetic flux densities, convection flow and continuum intensity
in the model granular cell. In Section 4, we discuss the
uncertainties resulted from the effect of projection. The last
section is devoted to conclusion.

\section{Observations and inversion strategy}

The SP observations in the SOT instruments aboard \emph{Hinode}
spacecraft provide the full Stokes spectral signals of two
magnetically sensitive FeI lines at 630.15 nm ($g_{eff}=1.67$) and
630.25 nm ($g_{eff}=2.5$). The SP performs four suitable
observations for scientific objectives: fast map, dynamics, normal
map and deep magnetogram. The data adopted here are the mode of
deep magnetogram because of its higher polarimetric
signal-to-noise ratio better than $10^{3}$.

Three quiet solar regions with the field-of-view of 55$''$.50
$\times$ 162$''$.30 were observed on 2007 June 1 (12:35 UT), June
2 ( 13:12 UT), and June 4 (01:20 UT), respectively. They are
located in the positions (-31$''$, -200$''$), (0$''$, -200$''$),
and (0$''$, 200$''$). The data in every region consist of 376
consecutive positions of spectrograph slit. For the observation of
each spectrograph slit, the exposure time is 9.6 seconds, yielding
a noise level in the polarization continuum of about $8.6\times
10^{-4}I_{c}$. The scanning steps of the spectrograph slit are
0$''$.1476, and the spectral sampling is 2.15 pm.

Vector magnetic fields are derived from the inversion of the full
Stokes profiles based on the assumption of Milne-Eddington
atmosphere model (Yokoyama 2008, in preparation). Inversion
techniques (e.g., Lites \& Skumanich 1990; Socas-Navarro 2001) are
robust when applied to the stronger Stokes polarization signals
from active regions, but these profiles-fitting procedures using
many free parameters typically encounter difficulties in
convergence toward and uniqueness of the solutions when confronted
with noisy profiles (Lites et al. 2008). However, Orozco Suarez et
al. (2007a) suggest that ME inversion turns out to be largely
independent of the noise and the field strength initialization,
provided they are running on these pixels showing polarization
signals above a reasonable threshold. Accordingly, here we only
analyze these pixels with total polarization degree above 3 times
of the noise level in the polarization continuum in order to
exclude some profiles which cannot be inverted reliably, and then
apply a `median' function to the inverted magnetogram for removing
the noisy peak.

The inversion returns the values of 13 free parameters, including
the three components of magnetic fields (the field strength $B$,
the inclination angle $\gamma $ with respect to the line-of-sight
(LOS) direction, the azimuth angle $\phi $), the stray light
fraction $\alpha$, the Doppler velocity $V_{Los}$, and so on.
Mart\'{i}nez Gonz\'{a}lez et al. (2006) demonstrate the pair of
FeI lines in low flux quiet Sun regions are not capable of
distinguishing between the intrinsic magnetic field and the
filling factor. Therefore, the flux density is a more appropriate
quantity to describe. Here, we show the equivalent, spatially
resolved vector magnetic fields by `apparent flux density' (Lites
et al. 1999) of the longitudinal and the transverse components,
that is $B^{L}_{app}=(1-\alpha)Bcos(\gamma)$ and
$B^{T}_{app}=(1-\alpha)^{\frac{1}{2}}Bsin(\gamma)$. The
longitudinal component $B^{L}_{app}$ may be thought of as the
magnitude of the LOS component of a spatially resolved magnetic
field that produces the observed circular polarization signal,
while the transverse component $B^{T}_{app}$ is vertical to the
LOS that would produce the observed linear polarization signal.
The noise level is 2 Mx $cm^{-2}$ for the longitudinal field, and
44 Mx $cm^{-2}$ for the transverse field.

Each spectrum requires 9.6 seconds, so a single 2$''$ granule may
be mapped in about 2 minutes, a shorter interval compared with the
evolution of granules (Bahng $\&$ Schwarzschild 1961; Mehltretter
1978; Title et al. 1986; Dialetis et al. 1986). Thusly, we assume
that we may use such a map to characterize the granules as if it
were a snapshot.

From the three quiet regions observed on 2007 June 1, 2, and 4,
totally 2,100 granules are selected manually according to the
image of continuum intensity, and there are distinct
inter-granular lanes surrounding each selected granule. Here, we
define the granular pattern including the surrounding
inter-granular lanes as granular cell, while the granular pattern
excluding the surrounding inter-granular lanes as granule. The
granular apparent flux densities, i.e., the mean unsigned
longitudinal and the mean transverse apparent flux densities
inside granular pattern excluding the inter-granular lanes, are
measured. Furthermore, the granular unsigned magnetic flux are
also calculated according to the granular area and the
corresponding unsigned longitudinal apparent flux density. In
addition, the granular intensity, Doppler velocity and magnetic
inclination, i.e., the mean intensity, the mean Doppler velocity
and the mean inclination within the granular pattern excluding the
inter-granular lanes, are also measured.

\section{The statistics of granular magnetic properties}

The Fig.1 shows the map of continuum intensity overlapped by the
longitudinal apparent flux density $B^{L}_{app}$ and transverse
apparent flux density $B^{T}_{app}$ for a small portion of the
observed region on June 1, 2007. In these small quiet regions,
about fourteen percents of the area represent the pixels which
have not been inverted because of their small polarization
signals.

The mean unsigned longitudinal apparent flux density in the
field-of-view of the integrated regions observed on 2007 June 1,
2, and 4 is 28 Mx $cm^{-2}$, with the corresponding mean
transverse apparent flux density of 98 Mx $cm^{-2}$. The mean
magnetic inclination to the surface that is vertical to the LOS
calculated over these inverted pixels is only 24.8 degrees. The
larger occurrence of transverse fields confirms the discovery of
strong linear polarization signals by Lites et al. (2007a, 2007b,
2008).

The current observations confirm the findings by Lites et al.
(2008) that the magnetic fields on the quiet Sun are not uniformly
distributed. Magnetic fields appear to be organized on
meso-granulation scales. Moreover, there are `\emph{voids}' in
either longitudinal or transverse fields where the flux density is
much lower than other locations, even close to zero flux density.
In this study we focus on the statistical properties of granular
vector magnetic fields and their overall correlation with
convection flow and intensity. Because the granular transverse and
unsigned longitudinal apparent flux densities are computed over
the whole granule, assigning zero fields to these pixels which are
not inverted, they represent the lower limits. However, the
granular magnetic inclinations are calculated over only these
pixels which are inverted. Considering the spatial resolution
0$''$.32 of the SOT/SP aboard the \emph{Hinode}, and the 1
$\sigma$ sensitivity of 2.0 Mx $cm^{-2}$ of longitudinal apparent
flux density, the minimum observable magnetic flux should be
$1.1\times10^{15}$Mx. Therefore, when the granular flux is less
than $1.1\times10^{15}$Mx, we think that the granular flux can not
be detectable.

The transverse apparent flux density, unsigned longitudinal
apparent flux density, Doppler velocity, and continuum intensity
of granules are measured. These physical parameters of 2,100
granules form a good sample. We refer these parameters as basic
MHD (magneto-hydrodynamics) parameters of granules. The
statistical MHD properties of the granules are listed in Table 1.

Here we define the granular relative intensity with respect to the
mean intensity of the quiet photosphere. We find that some
granules are darker than the mean photosphere in intensity. The
relative intensity of granules spans the range from 0.97 to 1.14
with the peak distribution at 1.07, just as shown in Fig.2.
Therefore, the granule is, of course, brighter than its
surrounding dark lanes, but not necessarily brighter than the mean
photosphere.

For studying the convection of granulation, the effects of solar
oscillations have to be separated from the velocity variations of
the granular pattern. In order to reduce the influence of the
oscillations, the method of `boxcar' smoothing function is adopted
(Krieg et al. 2000; Lites et al. 2008), i.e. applying a smoothing
function to the $V_{Los}$ image, then subtracting that image from
the $V_{Los}$ to obtain a `flattened' image $V_{Los}^{f}$ of the
local velocity variation which is mostly due to granulation. Here,
we smooth the $V_{Los}$ image in a width of 3$''$.2, and remove
the oscillation by subtracting that image from the $V_{Los}$ to
obtain the Doppler velocity related to the granular motion. The
distribution of granular Doppler velocity is shown in Fig.2. The
statistical analysis shows that the range of the granular motion
is from -3.3 km $s^{-1}$ (blueshift) to 2.0 km $s^{-1}$ (redshift)
with a peak at -1.0 km $s^{-1}$.

\subsection{The distributions of magnetic flux and apparent flux densities}

The magnetic flux distribution of the granules is presented in
Fig.3. Their flux spans the range of $1.1\times10^{15}$ Mx (the
minimum observable limit) to $3.3\times10^{18}$ Mx with a peak
distribution at $2.0\times10^{16}$ Mx, which is less than the peak
distribution at $6\times10^{16}$ Mx of inter-network magnetic
elements by the ground-based observations at poor spatial
resolution (Wang et al. 1995). This suggests that the magnetic
fields of granulation are likely to belong to the inter-network
magnetic fields. The mean granular flux is $1.4\times10^{17}$ Mx
(see Table 1). The peak distribution of granular area is 0.93
$Mm^{2}$, taking the mean unsigned longitudinal apparent flux
density of 28 Mx $cm^{-2}$ in the quiet region into account, and
the magnetic flux within the scale of granular size is
$2.6\times10^{17}$ Mx which is within the range of magnetic flux
of inter-network magnetic elements according to the mean flux
obtained by Wang et al. (1995).

The apparent flux density distributions are shown in Fig.4. It is
apparent from the figure that the unsigned longitudinal apparent
flux density of granules is weak with a mean apparent flux density
of 12 Mx $cm^{-2}$, and the granular transverse apparent flux
density spans the range from almost zero to 218 Mx $cm^{-2}$ with
a mean apparent flux density of 79 Mx $cm^{-2}$. Comparing with
the average apparent flux densities, 28 and 98 Mx $cm^{-2}$, of
the whole three regions, the magnetic fields in granules are
obviously weaker. This reflects a fact that the longitudinal and
transverse fields within the granule are weaker than those in the
inter-granular lanes. Another reason comes from the fact that our
statistics have excluded the abnormal granules which certainly
have much stronger apparent flux densities than the normal
granules. Moreover, the range of inclination with respect to the
surface that is perpendicular to LOS is from 4.8 degree to 76.7
degree with a peak distribution around 25 degree.

Considering the continuum intensity and inferred Doppler velocity,
the inter-granular lanes are separated from the granules. The mean
unsigned longitudinal and transverse apparent flux densities in
the inter-granular lanes are 29 Mx $cm^{-2}$ and 117 Mx $cm^{-2}$,
respectively, and the corresponding inclination to the surface
that is perpendicular to LOS only 20 degree. Therefore, in the
quiet region, the magnetic vectors of granule are more
\textbf{longitudinal} than those of inter-granular lanes. The
strong preference for longitudinal field resides in the
inter-granular lanes, which confirms the conclusion drawn by
Dom\'{i}nguez et al. (2003). Furthermore, the transverse field
also preferentially occurs in the inter-granular lanes.

\subsection{Correlations among the granular apparent magnetic flux density,
intensity and Doppler velocity}

The correlation between the granular Doppler velocity and
continuum intensity is studied. In order to show the relationship
clearly, we divide the granules into 100 equally populated bins
according to granular intensity, and compute the mean Doppler
velocity and the mean intensity in each bin, respectively. We find
that there is a negative correlation between Doppler velocity and
the continuum intensity, and their correlation coefficient is
-0.83, well above the confidence level of 99.9\%. We fit their
relationship by a linear function, exhibited in Fig.5.
\begin{equation}
V_{Los}=A_{0}+A_{1}\times I_{C}
\end{equation}
where $A_{0}=8.9\pm0.6$, $A_{1}=-9.0\pm0.6$. Therefore, we find
that the larger the spectral blueshift is, the brighter the
granule becomes.

In addition, by the same way, we also analyze the relationship
between the granular unsigned longitudinal apparent flux density
$B^{L}_{app}$ and the corresponding transverse apparent flux
density $B^{T}_{app}$. Our statistics show that there is a
positive correlation between them. We fit their relationship by
the polynomial function, just as displayed in Fig.6.
\begin{equation}
 B^{L}_{app}=A\times (B^{T}_{app})^{2}
\end{equation}
where $A=0.00164\pm0.00005$.

\

However, the statistics excludes the abnormal granulation because
the convective elements are modified by the stronger magnetic
field. In order to compare the properties of the abnormal granules
with that of the normal granules, we select arbitrarily three
abnormal granules manually in plage region. It is found that that
the mean unsigned longitudinal apparent flux density in the
abnormal granules are larger than 300 Mx $cm^{-2}$, while the
transverse apparent flux density are around 100 Mx $cm^{-2}$.
Their magnetic fluxes exceed $1.0\times10^{19}$ Mx, which supports
the conclusion of Cheung et al. (2007). Their maximum intensities
are higher than the average photospheric intensity by a factor of
1.1. Furthermore, due to the modification of the stronger magnetic
fields, the Doppler velocity of abnormal granules is very small.

\section{An average solar granular cell}

The variations of the apparent magnetic flux density, the Doppler
velocity, and the continuum intensity from the granular center to
the inter-granular lane are useful constraints for modelling the
solar granule. We construct a model of average solar granular
cell.

What we firstly done is to define the granular size. Generally
speaking, the granular size has two measures to define. One is the
distance between the centers of two adjacent granules; this size
is called the `cell size' (Stix 2002). Another is the granular
`diameter', defined as follows (see Bray et al. 1984): Let A be
the granular area over which a granular intensity is larger than a
certain value; the area generally is not circular, so an effective
diameter is $D=2(A/\pi)^{1/2}$. Using the 2100 granules manually
selected, we find that the peak distribution of granular
`diameter' D is 1$''$.50. In the image of continuum intensity in
an enlarged scale, we plot a horizontal line (from east to west)
and a vertical line (from south to north), and both of them span
the granular center and the corresponding inter-granular lanes. We
define the positions with the minimum intensity in inter-granular
lanes as the outer edge of a granular cell, and the distance
between the outer edges on both sides of the granular cell as the
size of granular cell $D_g$ which is larger than $D$ by the
definition. The $D_g$ is almost the same as the `cell size', the
distance between the centers of two adjacent granules.

Afterwards, we divide the $D_g$ into 26 equally populated bins
along the two lines mentioned above, and compute the mean
intensity in each bin. The measurements along the two orthogonal
lines are averaged to get the intensity changes from the granular
center to the corresponding inter-granular lanes. Making use of
the same method, we analyze one hundred granules which are, more
or less, regular with almost all of the Stokes profiles
well-inverted. Moreover, in order to reduce the asymmetry from
south-north cut, we select 50 granules in the quiet region (0
$''$, -200 $''$) and 50 granules in the quiet region (0 $''$, 200
$''$) to construct the average solar granular cell. Finally, we
average the intensity of one hundred granules in each bin, and
obtain the intensity variations with the distance from granular
center to its corresponding inter-granular lanes for an average
solar granular cell. In the image of continuum intensity, we can
see that there is a distinct boundaries, i.e., great contrast of
intensity, between granule and the corresponding inter-granular
lanes. Therefore, from granular center to inter-granular lanes, we
can also find two adjacent bins with largest contrast of
intensity, and define the bin with less intensity as the inner
edge of the average solar granular cell. We define the distance
between the inner edges as the effective diameter $D$ of the
average solar granule, and this distance is bounded by the two
dashed lines in fig.7 . For the average solar granule, the
diameter $D$ is equal to 1$''$.50 according to the peak
distribution of granular effective diameter, and the size $D_g$ of
the granular cell is 2$''$.16 based on the ratio between $D$ and
$D_g$ of the average solar granular cell.

Finally, by the same ways we obtain the average Doppler velocity,
unsigned longitudinal and transverse magnetic flux densities of
one hundred granules in each bin; and the variations of apparent
magnetic flux densities and Doppler velocity with the distance
from granular center are obtained. The detailed distribution of
apparent flux densities, convective flow and intensity for the
average solar granular cell is schematically drawn in Fig.7. It is
interesting to notice that the MHD parameter distributions of
average granular cell are not exactly symmetric. Furthermore, we
find that there is also asymmetry when we use only the results
from the east-west cuts for the average solar granular cell.

For the average solar granular cell which we construct, the mean
unsigned longitudinal and transverse apparent flux densities in
inter-granular lanes are 22 Mx $cm^{-2}$ and 102 Mx $cm^{-2}$,
respectively; whereas that in granule, only 7 Mx $cm^{-2}$ and 61
Mx $cm^{-2}$. The apparent flux density distributions of the
average solar granular cell confirm the previous results that
there is a preference for longitudinal and transverse fields to
reside in the inter-granular lanes. The magnetic flux of the
surrounding dark lanes of the average solar granular cell is
$2.2\times 10^{17}$ Mx, while the magnetic flux of the average
solar granule excluding the dark lanes is only $6.5\times 10^{16}$
Mx.

From Fig. 7, we see clearly that the plasma motion corresponding
to the LOS in the granular center is fastest, and there is a
negative correlation between continuum intensity and Doppler
velocity. The maximum intensity contrast of the average solar
granule is $C=2\frac{I_{max}-I_{min}}{I_{max}+I_{min}}=18\%$
(where $I_{max}$ and $I_{min}$ are the maximum and minimum of the
intensity). The effective diameter, $D$, of average granule is
larger than the width occupied by its inter-granular lanes by a
factor of 2.27.

If we define the radius $D_g/2$ of granular cell as unit one, then
the apparent flux densities can be expressed as the function of
the distance from granular center, $d$. The relationship between
$B^{L}_{app}$ and distance $d$ can be fitted by a two-peak
Gaussian function, displayed in Fig.8.
\begin{equation}
  B^{L}_{app}=B_{L0}+\frac{B_{L1}}{\sqrt{\frac{\pi}{2}}\sigma_{11}}e^{-2(\frac{d-\mu_{11}}{\sigma_{11}})^2}
  +\frac{B_{L2}}{\sqrt{\frac{\pi}{2}}\sigma_{12}}e^{-2(\frac{d-\mu_{12}}{\sigma_{12}})^2}
\end{equation}
where $B_{L0}=33.9\pm0.8$, $B_{L1}=-17.9\pm0.6$,
$\sigma_{11}=0.66\pm0.03$, $\mu_{11}=- 0.41\pm0.02$,
$B_{L2}=-28.4\pm0.6$, $\sigma_{12}=0.84\pm0.03$,
$\mu_{12}=0.37\pm0.02$. The effective half-width of $B^{L}_{app}$
is 0.765.

The relationship between $B^{T}_{app}$ and granular diameter d can
be fitted by a single Gauss function, just as shown in Fig.9.
\begin{equation}
  B^{T}_{app}=B_{T0}+\frac{B_{T1}}{\sqrt{\frac{\pi}{2}}\sigma_{2}}e^{-2(\frac{d-\mu_{2}}{\sigma_{2}})^2}
\end{equation}
where $B_{T0}=158.5\pm0.3$, $B_{T1}=-200\pm2$,
$\sigma_{2}=1.5\pm0.1$, $\mu_{2}=0.005\pm0.005$. The half-width of
$B_{T}$ is 0.73.

In addition, the Doppler velocity $V_{Los}$ and continuum
intensity $I_{C}$ also can be expressed as the function of $d$ by
the two-peak Gaussian fit.
\begin{equation}
  V_{Los}=V_{0}+\frac{V_{1}}{\sqrt{\frac{\pi}{2}}\sigma_{31}}e^{-2(\frac{d-\mu_{31}}{\sigma_{31}})^2}
  +\frac{V_{2}}{\sqrt{\frac{\pi}{2}}\sigma_{32}}e^{-2(\frac{d-\mu_{32}}{\sigma_{32}})^2}
\end{equation}
where $V_{0}=1.26\pm0.07$, $V_{1}=-1.7\pm0.2$,
$\sigma_{31}=0.67\pm0.04$, $\mu_{31}=0.39\pm0.04$,
$V_{2}=-1.9\pm0.3$, $\sigma_{32}=0.73\pm0.05$,
$\mu_{32}=-0.28\pm0.04$.
\begin{equation}
  I_{C}=I_{0}+\frac{I_{1}}{\sqrt{\frac{\pi}{2}}\sigma_{41}}e^{-2(\frac{d-\mu_{41}}{\sigma_{41}})^2}
  +\frac{I_{2}}{\sqrt{\frac{\pi}{2}}\sigma_{42}}e^{-2(\frac{d-\mu_{42}}{\sigma_{42}})^2}
\end{equation}
where $I_{0}=0.887\pm0.004$, $I_{1}=0.16\pm0.01$,
$\sigma_{41}=0.74\pm0.03$, $\mu_{41}=- 0.33\pm0.02$,
$I_{2}=0.16\pm0.01$, $\sigma_{42}=0.73\pm0.02$,
$\mu_{42}=0.37\pm0.02$.

\section{Discussions}

So far, we only use the terms of longitudinal and transverse
apparent flux densities to describe the results of our statistics.
Because the three quiet regions are not observed at disk center,
there will appear some uncertainties resulted from the effect of
projection in the estimation of vertical and horizontal fields.
Accordingly, the longitudinal and transverse apparent flux
densities do not strictly mean the vertical and horizontal
apparent flux densities. Thusly, we make a simple estimation about
the uncertainties. We find that the uncertainty is about 4 Mx
$cm^{-2}$ for the vertical field, and 7 Mx $cm^{-2}$ for the
horizontal field. The uncertainties are within 3 times of the
standard deviations of magnetic flux density measurements.
Therefore, it is quite acceptable to use the terms of vertical and
horizontal fields to describe our statistical results on the
vector magnetic fields of solar granulation. Compared with the
values of magnetic parameters from 2100 granules and average solar
granular cell, we find that the vertical fields are weaker than
the corresponding horizontal field in the whole quiet region. The
vertical fields are concentrated in the inter-granular lanes,
which confirmed the conclusion about the distribution of vertical
fields obtained by Lites et al. (2008). Moreover, we find that the
horizontal fields also occur most commonly in the inter-granular
lanes. This conclusion is different from the result about the
horizontal fields distribution obtained by Lites et al. (2008).
The reason for the difference may be the fact that the result
showed by Lites et al. (2008) mostly emphasizes the fraction of
the area of the bin occupied by the greater horizontal field for
equally populated bins from the granular center to inter-granular
lanes, while our result stresses the magnitude of horizontal field
for each bin from the granular center to inter-granular lanes.

On the other hand, the effect of projection could also result in
the uncertainties of less than 12 degrees in the estimation of
magnetic inclination angle with respect to the horizontal solar
surface. Compared with the magnetic inclination angles with
respect to the image surface being perpendicular to the
line-of-sight, 27.1 degree and 20 degree, for the granules and
inter-granular lanes, the uncertainty of magnetic inclination
angle makes us hardly affirm that the magnetic fields in the
granules are more vertical than that in the inter-granular lanes.
Therefore, we select a quiet region observed at disk center on
2007 December 21, and divided the region into granular region and
inter-granular region according to the continuum intensity and the
inferred Doppler velocity. The magnetic inclination angles are
averaged over only the inverted pixels in the granular region and
inter-granular region, respectively. We find that average magnetic
inclination angle with respect to line-of-sight is 53.2 degrees in
granules, and 59.2 degrees in the inter-granular lanes, i.e., 37.8
and 30.8 degrees with respect to the local solar surface.
Therefore, we confirm that the magnetic fields in the granules are
more vertical than that in the inter-granular lanes. However,
Orozco Su$\acute{a}$rez et al. (2007b) show that the fields are
more horizontal in granular region than elsewhere seen from the
probability density functions of magnetic inclination, which is
opposite to our result. Indeed, more studies should be made to
clarify the discrepancy.

\section{Conclusions}

Based on the observations of quiet Sun using SOT/SP aboard the
\emph{Hinode} spacecraft, we have statistically studied the
distributions of granular magnetic flux, vector magnetic fields,
the continuum intensity and the Doppler velocity. Moreover, a
model of average solar granular cell is reconstructed.

The granular flux spans the range from $1.1\times 10^{15}$ Mx to
$3.3\times 10^{18}$ Mx with a peak distribution at $1.6\times
10^{16}$ Mx. The granular vertical apparent flux density is weaker
than the corresponding horizontal apparent flux density, and there
is a positive correlation between them. The range of granular
inclination to the surface that is vertical to the LOS is from 4.8
to 76.7 degrees with a peak distribution at 25 degrees, and the
magnetic vectors in the granules is more vertical than that in the
inter-granular lanes. There is a strong preference that the
vertical and horizontal fields reside in the inter-granular lanes.
There is a range for the Doppler velocity from -3.3 km $s^{-1}$
(blueshift) to 2.0 km $s^{-1}$ (redshift) after removing the
velocity components due to the oscillation. Some granules are
darker than the quiet photosphere in intensity, the relative
intensity of granules has a range spanning from 0.97 to 1.14.
There is a linear correlation between granular intensity and
Doppler velocity.

A model of an average solar granular cell has been built by
analyzing one hundred granules, and the variations of the vector
magnetic fields, the Doppler velocity and the continuum intensity
from the inter-granular lanes to the granular center are studied.
For the average solar granular cell, it is also confirmed that the
vertical and horizontal fields have a strong preference in the
inter-granular lanes. The changes of apparent flux densities,
convection flow, and relative intensity with the distance from
granular center to the corresponding inter-granular lanes can be
empirically formulated. Interestingly, the transverse apparent
flux density distribution can well be formulated by a Gaussian
function; however, the unsigned longitudinal field distributions
can only be best fitted by double-peak Gaussian functions. These
empirical formulation may serve as some quantitative constraints
on the magneto-convection simulations.

 \acknowledgments

The authors are grateful to the \emph{Hinode} team for providing
the data. \emph{Hinode} is a Japanese mission developed and
launched by ISAS/JAXA, with NAOJ as a domestic partner and NASA
and STFC (UK) as international partners. It is operated by these
agencies in co-operation with ESA and NSC (Norway). In addition,
the authors are also thankful to Dr. Guiping Zhou of National
Astronomical Observatories for her helpful suggestions. This work
is supported by the National Natural Science Foundations of China
(10703007, G10573025, 40674081 and 10603008), the CAS Project
KJCX2-YW-T04, and the National Basic Research Program of China
(G2006CB806303).

\begin{table}[htbp]
\begin{center}
\caption{The properties of granulation. \label{tbl-2}}
\begin{tabular}{crrrrrrrrrrr}
\tableline\tableline
 Term       &  average   &  range  \\
\tableline
Continuum intensity(n-dim)..................& 1.07  & 0.97$--$1.14\\
Doppler velocity(km $s^{-1}$)......................& -0.82     & -3.3$--$2.0\\
Total magnetic flux($10^{16}$Mx)..............& 14.2  &  0.1$--$334.4\\
Unsigned longitudinal apparent flux density(Mx $cm^{-2}$) .........& 12    &0$--$212\\
Transverse apparent flux density(Mx $cm^{-2}$)..........................& 79      &4$--$218\\
Inclination to the surface being vertical to the LOS(degrees)...   & 27.1    &4.8$--$76.7\\
\tableline\tableline
\end{tabular}
\end{center}
\end{table}
\clearpage

\begin{figure}
\epsscale{1.4} \plotone{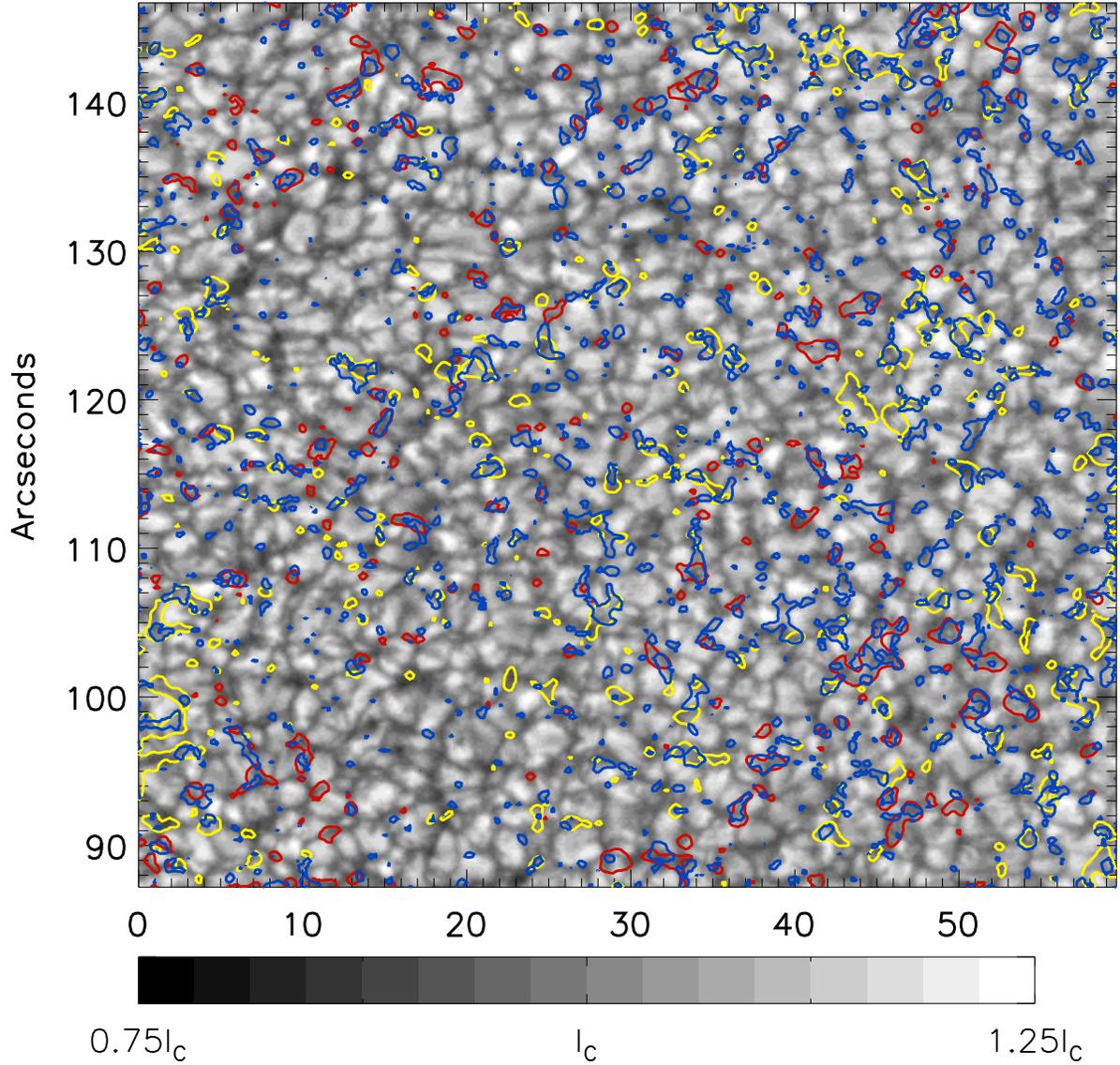} \caption{Continuum intensity with
the contours of the transverse and longitudinal apparent flux
densities. Yellow and red contours show, respectively, the
negative and positive 50 Mx $cm^{-2}$ longitudinal apparent flux
density levels. The blue contour are 180 Mx $cm^{-2}$ level for
the transverse apparent flux density.
  \label{fig1}}
\end{figure}


\begin{figure}
\epsscale{1.0} \plotone{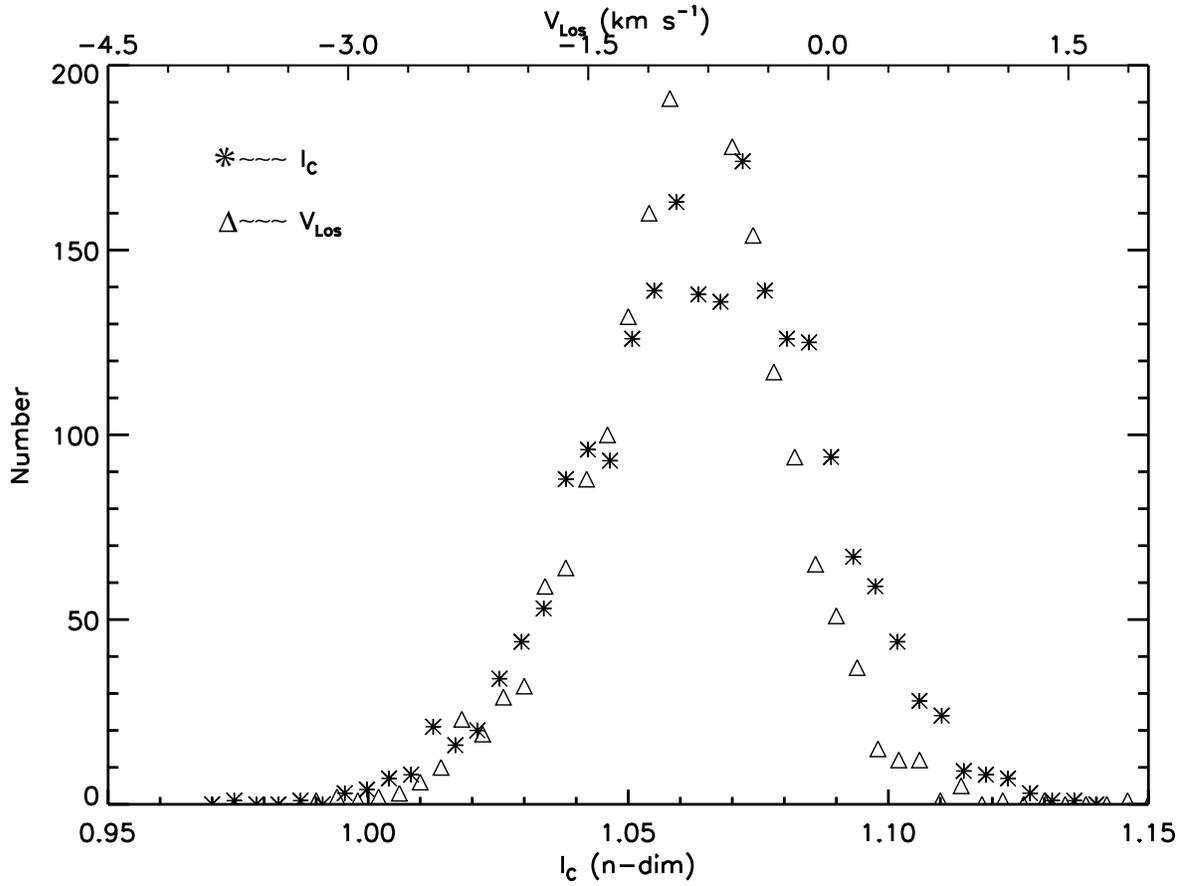} \caption{The distributions of
Doppler velocity displayed by triangle and continuum intensity
shown by asterisk. The Doppler velocity is shown on the top scale,
while the continuum intensity on the bottom scale. The continuum
intensity is the relative intensity to the mean photosphere.
  \label{fig1}}
\end{figure}

\begin{figure}
\epsscale{1.0} \plotone{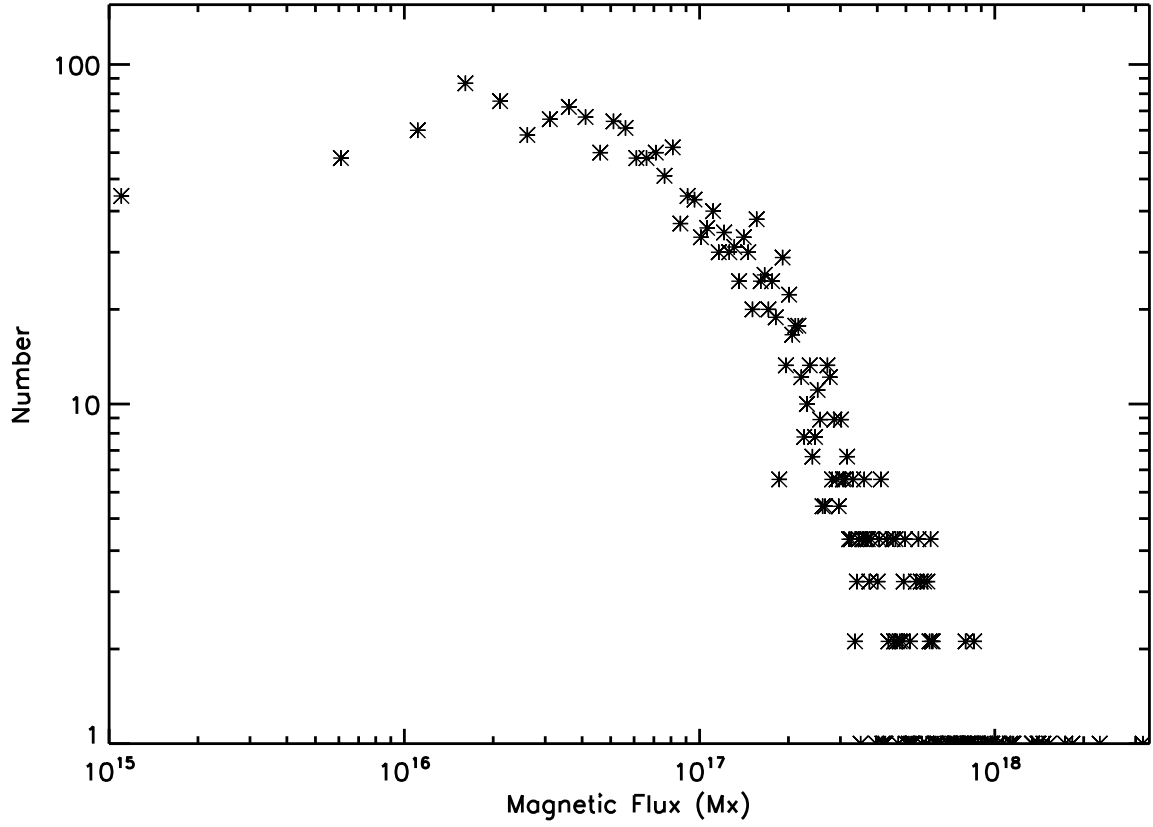} \caption{The unsigned magnetic
flux distribution of the granules.
  \label{fig1}}
\end{figure}

\begin{figure}
\epsscale{1.0} \plotone{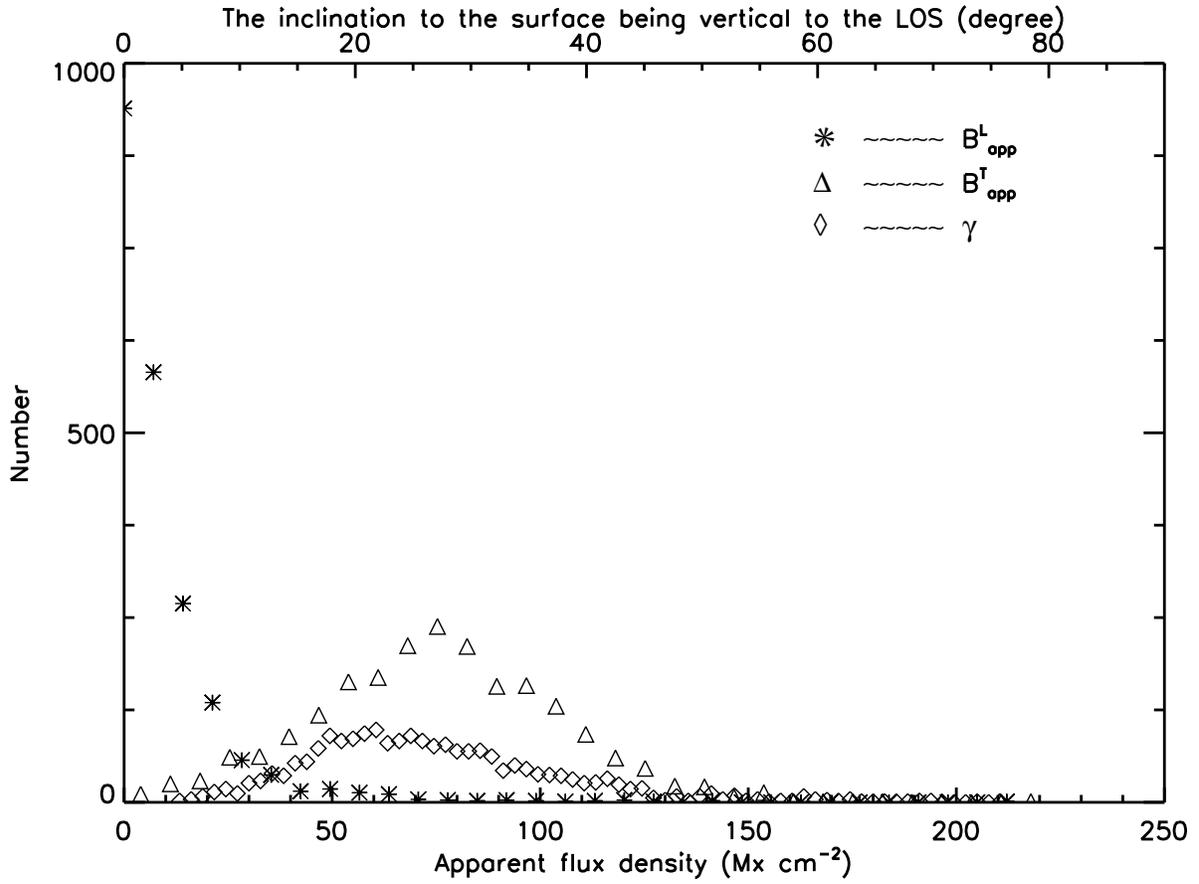} \caption{The distributions of
granular apparent flux densities. The unsigned longitudinal
apparent flux density denoted by asterisk, and the transverse one
shown by triangle, adopt a common scale (bottom scale), while the
distribution of the inclination to the surface being vertical to
the LOS displayed by diamond on the top scale.
  \label{fig1}}
\end{figure}

\begin{figure}
\epsscale{1.0} \plotone{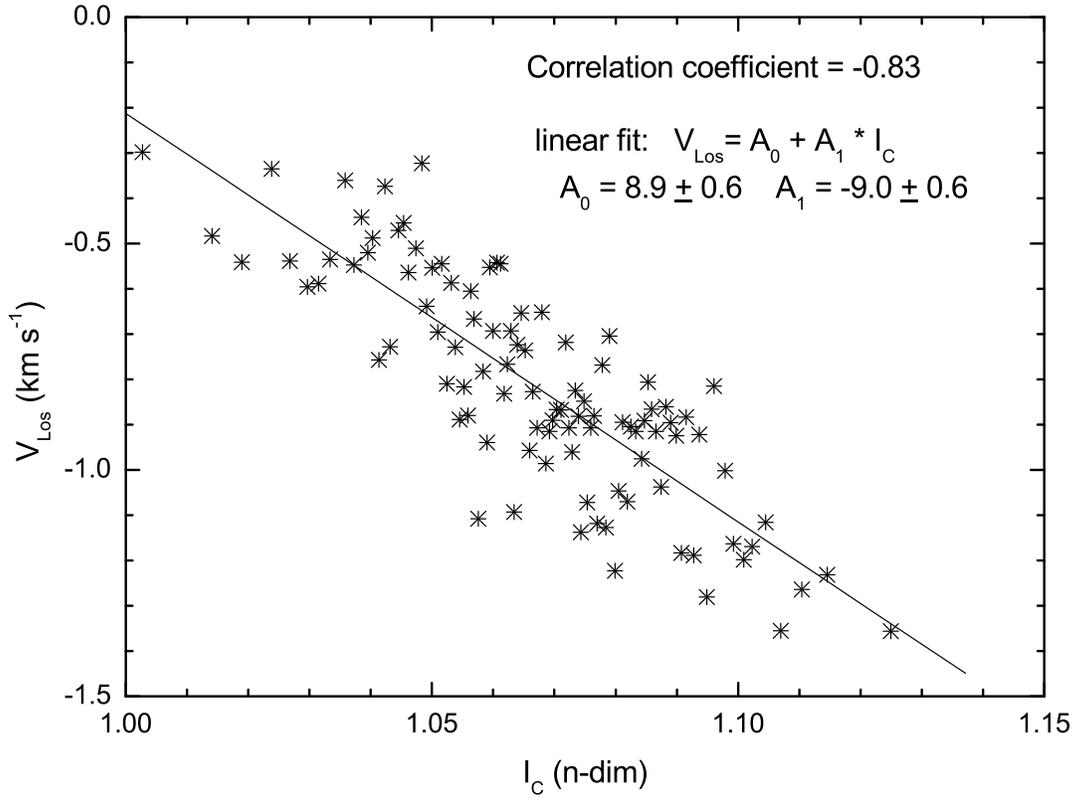} \caption{The relationship between
the granular relative intensity (non-dimension) and the granular
Doppler velocity (km $s^{-1}$). In order to show their
relationship clearly, the granules are divided into 100
equal-population bins according to the magnitude of relative
intensity, and the mean intensity and mean velocity in each bin
are calculated. The solid line shows the linear fit on the
granular intensity and Doppler velocity
  \label{fig1}}
\end{figure}

\begin{figure}
\epsscale{1.0} \plotone{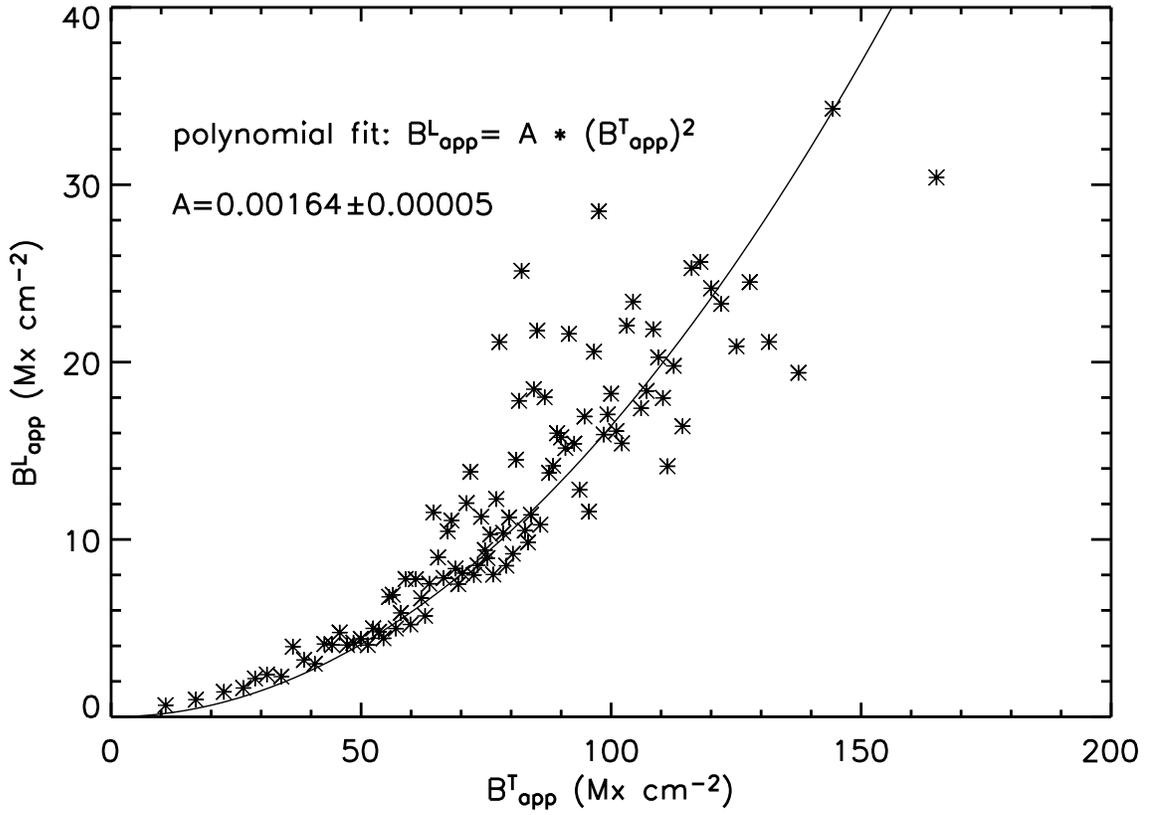} \caption{The relationship between
the granular transverse apparent flux density, $B^{T}_{app}$, and
the unsigned longitudinal apparent flux density, $B^{L}_{app}$.
The solid line represents the linear fit. \label{fig1}}
\end{figure}

\begin{figure}
\epsscale{1.0} \plotone{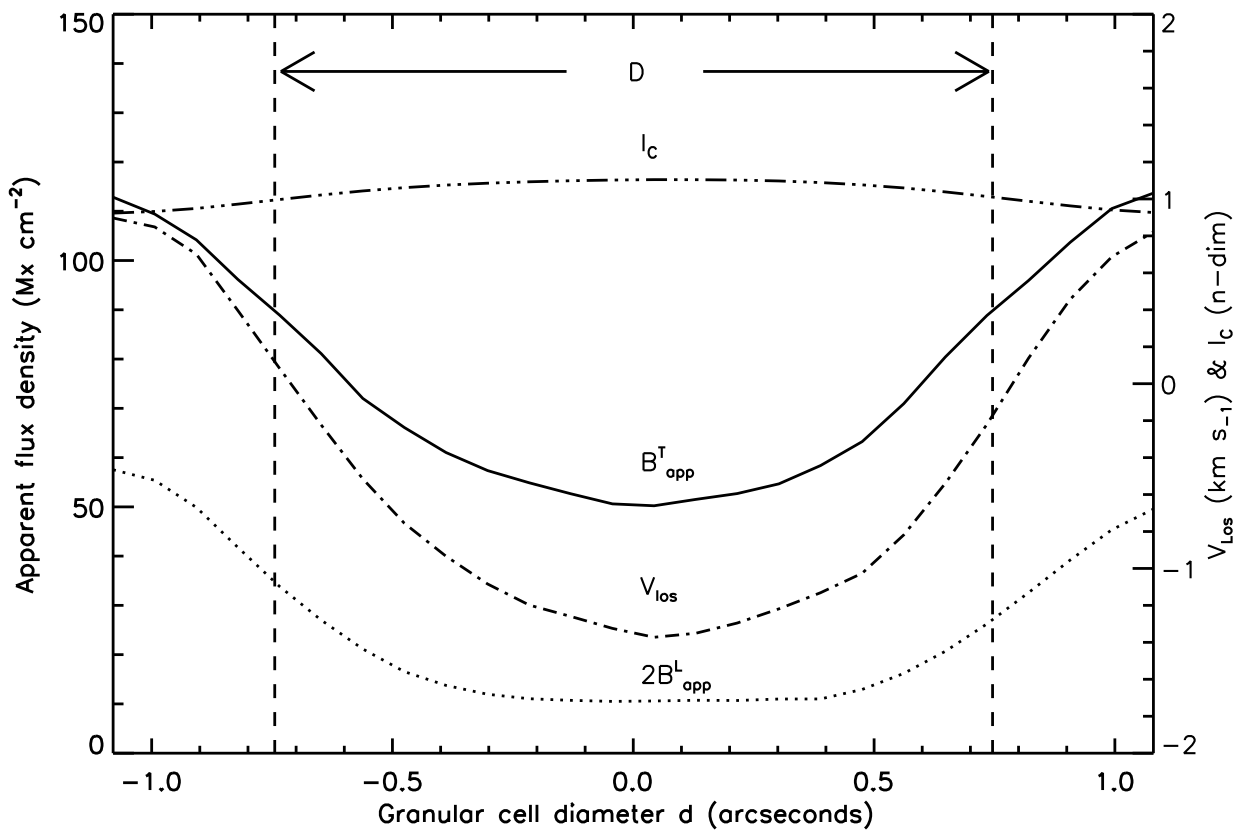} \caption{A model of average solar
granular element. Note that the apparent flux densities (solid
line and dotted line) are displayed on the left scale, while the
Doppler velocity (dash dot line) and the relative intensity (dash
dot dot line) on the right scale. The unsigned longitudinal
apparent flux density has been scaled by a factor of 2. The two
vertical dashed lines represent the edge between granule and
inter-granular lanes, and $D$ is the effective diameter.
  \label{fig1}}
\end{figure}

\begin{figure}
\epsscale{1.0} \plotone{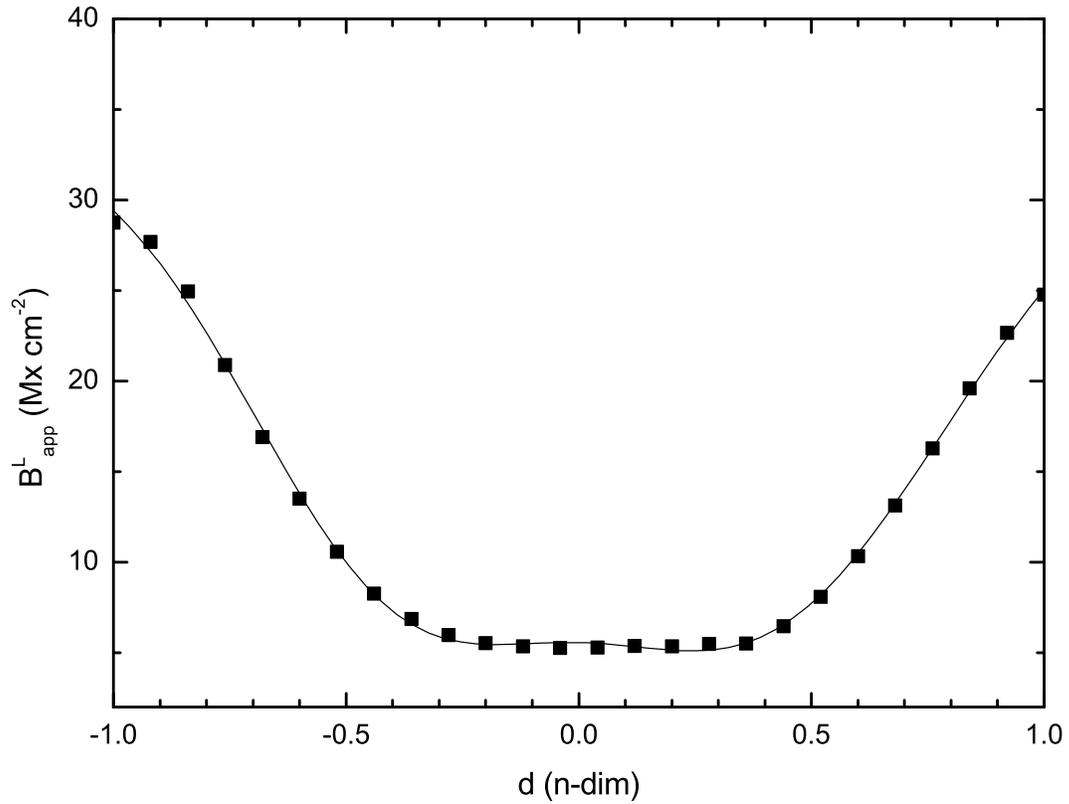} \caption{The variation on unsigned
longitudinal apparent flux density with granular diameter in
average solar granular element. The squares display the
longitudinal apparent flux densities of the reconstructed average
solar granular element. The solid line shows the multi-peaks Gauss
fit. The radius $D_g/2$ is defined as unit one.
  \label{fig1}}
\end{figure}

\begin{figure}
\epsscale{1.0} \plotone{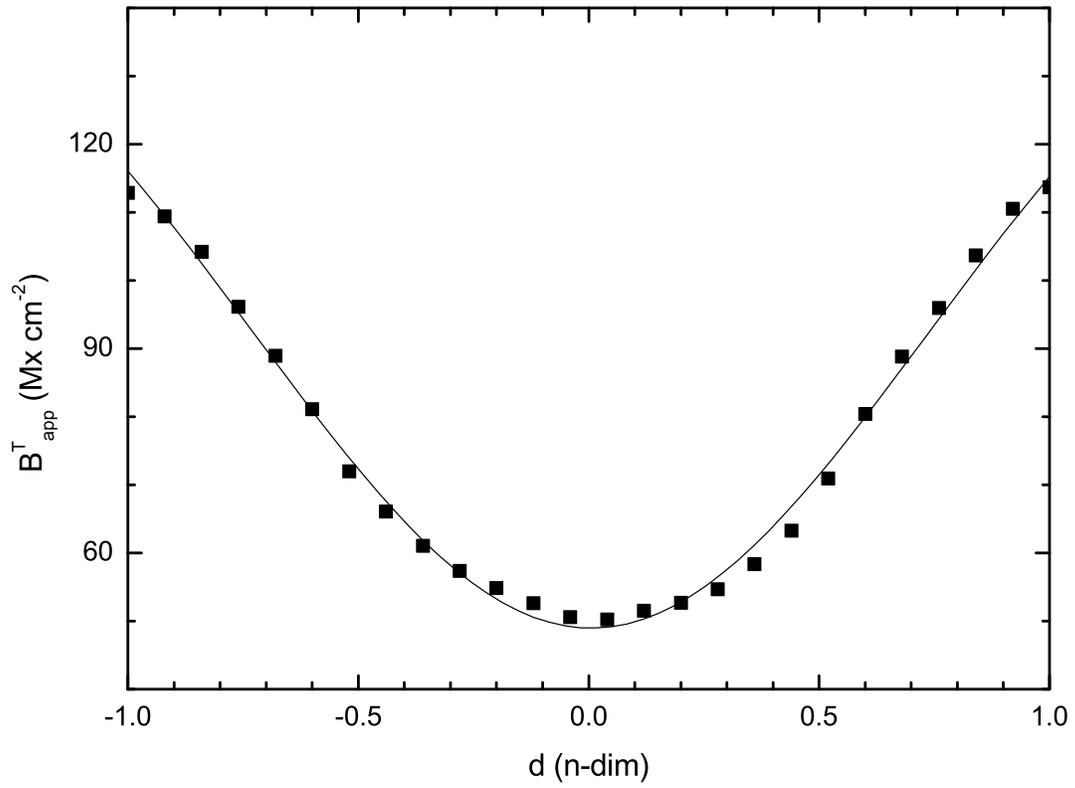} \caption{The variation on
transverse apparent flux density with the granular diameter in
average solar granular element. The solid line shows the Gauss fit
function on transverse apparent flux density variation.
  \label{fig1}}
\end{figure}

\end{document}